\documentstyle[12pt,epsfig]{article}  % impleLaTeX, use with DINA4 as follows  
\newlength{\dinwidth}                       
\newlength{\dinmargin}                      
\def\lsim{\mathrel{\rlap{\lower4pt\hbox{\hskip1pt$\sim$}}
    \raise1pt\hbox{$<$}}}                % less than or approx. symbol
\def\gsim{\mathrel{\rlap{\lower4pt\hbox{\hskip1pt$\sim$}}
    \raise1pt\hbox{$>$}}}                % greater than or approx. symbol
\def\beq{\begin{equation}}
\def\eeq{\end{equation}}
\def\beqn{\begin{eqnarray}}
\def\eeqn{\end{eqnarray}}
\def\pl#1#2#3{{\it Phys. Lett. }{\bf #1}(19#2)#3}
\def\zp#1#2#3{{\it Z. Phys. }{\bf #1}(19#2)#3}
\def\prl#1#2#3{{\it Phys. Rev. Lett. }{\bf #1}(19#2)#3}

\def\pr#1#2#3{{\it Phys. Rev. }{\bf #1}(19#2)#3}
\def\np#1#2#3{{\it Nucl. Phys. }{\bf #1}(19#2)#3}

\def\nim#1#2#3{{\it Nucl.  Instr. \& Meth. }{\bf #1}(19#2)#3}

\def\sss{\scriptscriptstyle}
\def\as{\alpha_{\sss S}}         
\begin{document}

{\flushright{
        \begin{minipage}{4cm}
        ETH-TH/96-34 \hfill \\
        hep-ph/yymmxxx\hfill \\
        \end{minipage}        }

}

\begin{center}  \begin{Large} \begin{bf}
Heavy Flavour Production\footnote{~To appear in the proceedings 
of the workshop {\it Future Physics at HERA}, eds. G.~Ingelman, 
A.~De Roeck and R.~Klanner, DESY, Hamburg, 1996.}\\
  \end{bf}  \end{Large}
  \vspace*{5mm}
  \begin{large}
R. Eichler, S. Frixione\\
\end{large}
\vspace*{0.3cm}
ETH-Z\"urich, CH 8093 Z\"urich, Switzerland \\
\end{center}

\begin{quotation}
\noindent
{\bf Abstract:}
We discuss the impact of the planned upgrades of the HERA collider
on the study of open heavy flavour and quarkonium production.
New experimental techniques in charm physics are presented.
\end{quotation}

\section{Introduction}

The $ep$ collider HERA offers new opportunities to study the
production mechanism of heavy quarks and to test the predictions
of the underlying theory. In the following, we will deal
with the processes
\beqn
e\,+\,p\,\,&\longrightarrow&\,\,Q\,+\,\overline{Q}\,+\,X,
\label{electroproduction}
\\
\gamma\,+\,p\,\,&\longrightarrow&\,\,Q\,+\,\overline{Q}\,+\,X,
\label{photoproduction}
\\
\gamma\,+\,p\,\,&\longrightarrow&\,\,\left(Q\overline{Q}\right)\,+\,X,
\label{photoquarkonia}
\eeqn
where $\left(Q\overline{Q}\right)$ indicates a generic heavy quarkonium
state, like $J/\psi$. The dominant contribution to 
the $Q\overline{Q}$ cross section is due to those
events in which the virtuality of the photon exchanged between
the electron and the proton is very small. In this case, the electron 
can be considered to be equivalent to a beam of on-shell photons, whose
distribution in energy (Weizs\"acker-Williams function~\cite{WWfunction})
can be calculated in QED. The underlying production mechanism
is therefore a photoproduction one (eqs.~(\ref{photoproduction})
and~(\ref{photoquarkonia})), which has been studied extensively 
in fixed-target experiments.
At HERA, the available center-of-mass energy is about one
order of magnitude larger than at fixed-target facilities; 
this energy regime is totally unexplored 
in photoproduction, and several new features have
to be taken into proper account. In particular, the large contribution
of the resolved photon component introduces in the theoretical
predictions a source of uncertainty which is totally negligible
at fixed-target energies.

A complementary way of studying heavy flavour production at HERA
is to retain only those events characterized by a large photon
virtuality (eq.~(\ref{electroproduction})). Although the total 
rates are much smaller than the photoproduction ones, the resolved
component is completely eliminated and more reliable theoretical
and experimental results can be obtained. Also, the dependence
of the data upon the photon virtuality can be used as a further
test of QCD predictions.

\section{Theoretical summary}

Fully exclusive, next-to-leading order perturbative QCD 
calculations~\cite{NLOcalc,NLO_ep,NLO_epdiff} are now 
available for the processes 
of eqs.~(\ref{electroproduction}) and~(\ref{photoproduction}).
In the framework of the factorization theorem of ref.~\cite{Bodwin95},
a next-to-leading order calculation for the direct color-singlet
photoproduction of quarkonium has been presented in 
ref.~\cite{Kraemer95}. Other contributions to the production of quarkonium,
like the resolved photoproduction and the direct color-octet photoproduction,
which are potentially relevant in kinematical regions accessible
at HERA, have been considered at the leading order in 
ref.~\cite{COquarkonia}.

In spite of these substantial progresses in understanding the
production mechanism, the (fixed-order perturbative) results
may become unreliable in certain kinematical regions, due to the 
appearance of potentially large logarithms which spoil the convergence 
of the perturbative expansion. In this case, a resummation to all orders 
of these large logarithms has to be performed.

When the available center-of-mass energy $\sqrt{S}$ gets large, the 
effective expansion parameter of the perturbative series becomes
\mbox{$\as\log(S/m_Q^2)$}. The problem of resumming these terms
({\it small}-$x$ {\it effects}) has been tackled by several 
authors~\cite{res_smallx}, mainly in the context of $b$ production 
in hadronic collisions. Specific studies for HERA~\cite{smallx_HERA} 
lead to the conclusion that the total photoproduction rates can be increased
by the resummation of the $\log(S/m_Q^2)$ terms by a factor smaller than
or equal to 40\% (45\%) with respect to the next-to-leading
order prediction for the direct (resolved) contribution in charm
production. In the case of bottom, the effect is much milder,
being always smaller than 10\%.

The transverse momentum distribution is in principle affected 
by the presence of~\mbox{$\log(p_{\sss T}/m_Q)$} terms. These logarithms
can be resummed by observing that, at high $p_{\sss T}$, the heavy-quark
mass is negligible, and by using perturbative fragmentation 
functions~\cite{Cacciari96}. Remarkably enough, the fixed-order and
the resummed results agree in a very wide range in $p_{\sss T}$ for
charm production (in the HERA energy range, the contribution of the 
resummation of this kind of logarithms is expected to be negligible for 
bottom production); the effect of the resummation might be visible only
for $p_{\sss T}$ larger than 50 GeV.

Finally, multiple soft gluon emission makes the perturbative expansion
unreliable close to the threshold or to the borders of
the phase space, like for example the regions 
$p_{\sss T}^{Q\bar{Q}}\simeq 0$ and $\Delta\phi^{Q\bar{Q}}\simeq \pi$.
This problem has not been directly dealt with during the workshop, but 
a lot of theoretical work has recently been performed~\cite{res_softgl}.
The resummation of soft gluons would be of great help
in order to have a reliable prediction for $J/\psi$ production
at $p_{\sss T}\simeq 0$; also, threshold effects are important
for $b$ production at HERA-B.

\subsection{Quarkonium Photoproduction}

Quarkonium photoproduction has been dealt with in ref.~\cite{Cacciari_proc}.
The benchmark process is in this case the production of $J/\psi$.
With suitable cuts, typically $p_{\sss T}>1$~GeV and $z<0.9$, where
\beq
z\,=\,\frac{k(p)\cdot k(J/\psi)}{k(p)\cdot k(\gamma)},
\eeq
the dominant contribution to the cross section is due to the direct
color-singlet production of $J/\psi$; this cross section has been
calculated at next-to-leading order in QCD~\cite{Kraemer95}.
The radiative corrections turn out to be sizeable in the HERA
energy range, being of the same order of the leading-order
contribution. The total rate is of the order of tens of $nb$,
and a first comparison with data~\cite{Aid96} has already been performed.
Although there is a good agreement as far as the shape of
($z$ and $p_{\sss T}$) distributions is concerned, the theoretical
predictions appear to slightly undershoot the data for the
total cross section; this can be adjusted by properly tuning the
input parameters of the calculation, since the normalization
is affected by large uncertainties. With a larger integrated
luminosity, a more significant comparison will be possible,
eventually helping in constraining the input parameters.

The luminosity upgrade is mandatory in order to assess the importance
of the $J/\psi$ production mechanisms neglected so far, namely
the direct color-octet contribution and the resolved contribution.
The former has been considered in refs.~\cite{COquarkonia,Cacciari_proc}
at the leading order in QCD. It has to be stressed that this contribution
can not be calculated from first principles, and relies upon
fits to Tevatron data on color-octet $J/\psi$ production. The
comparison with HERA data can therefore be regarded as a test of
the underlying factorization picture~\cite{Bodwin95}. Using the
result of ref.~\cite{COquarkonia}, the color-octet contribution
appears to be dominant over the color-singlet one
in the large-$z$ region; the first HERA data
do not support this behaviour. Nevertheless, since both the data
and the theoretical calculations need some refinement, no
definite conclusion can be drawn at present.
The resolved $J/\psi$ production is expected to be dominant
at very low-$z$ values, being enhanced by the color-octet terms.
Since the total rate is ${\cal O}(1~nb)$ in that region,
large integrated luminosity is needed to study this problem.

To further investigate the interplay between color-singlet and
color-octet contributions, other possibilities were taken
into account in ref.~\cite{Cacciari_proc}. For example, the
photoproduction cross section of $\chi$ should be dominated
by color-octet terms; the total rate is however suppressed
by two orders of magnitude with respect to the one of $J/\psi$,
and therefore large luminosity is required. The $J/\psi +\gamma$
production has a very distinctive signature to color-octet
contribution in the large-$z$ region~\cite{CGK}, but the cross
section is very small (${\cal O}(10~pb)$). Finally, large
luminosity is necessary also in order to investigate $J/\psi$
production for $p_{\sss T}>10$~GeV, where the dominant contribution
is expected to be due to charm fragmentation (the total rate is
at most ${\cal O}(1~pb)$).

\subsection{Open Heavy Flavour Production}

Several aspects of the deep-inelastic production of heavy flavours at HERA 
have been studied during the workshop. The work of ref.~\cite{Laenen_proc}
basically falls into three parts.
To begin with, the authors present a study of the heavy quark
inclusive structure function $F_2(x,Q^2,m^2)$, its sensitivity
to the input parameters of the calculation, and the size of the 
QCD corrections. The full next-to-leading order theoretical
prediction~\cite{NLO_ep,NLO_epdiff} suffers from a fairly little 
uncertainty, and the radiative corrections are not too large.
The structure function is on the other hand significantly sensitive to
the small-$x$ behaviour of the gluon density in the proton. 
Therefore, it is concluded that $F_2$ in charm production is an excellent 
probe to infer the gluon density in the proton at small $x$. 

Next, the exclusive properties of the final state are investigated,
since they allow a deeper understanding of the dynamics of the production 
process with respect to the inclusive structure function. 
The next-to-leading order corrections to $F_k,\;k=2,L$, in a fully
differential form have been recently calculated~\cite{NLO_epdiff} using 
the subtraction method. The results were incorporated in a 
Monte-Carlo style program which allows one to study correlations 
in the laboratory frame. The authors of ref.~\cite{Laenen_proc} 
present some single-inclusive distributions and
compare them with preliminary ZEUS data~\cite{ZEUS_warsaw}. 
In general, the data follow the shape of the theoretical curves,
but lie above them. This behaviour is consistent with the recent
analysis by H1~\cite{Adloff96} that, using a single-inclusive quantity,
showed that the charm production mechanism is boson-gluon fusion, 
and not a consequence of charm quarks in the sea.
Correlations between charm and anticharm are also presented.
Clearly, all these issues will be further clarified as soon as data
with better statistics will be available; a large integrated
luminosity is highly desirable.

Finally, the authors of ref.~\cite{Laenen_proc} deal with the
fact that, at truly large $Q^2$, a charm quark should be described 
as a light quark, i.e. as a constituent parton of the proton, whereas 
at $Q^2$ of order of $m^2$ it can only be produced through boson-gluon fusion
(see ref.~\cite{charmparton} for a detailed discussion).
It is shown that, surprisingly, already at $Q^2$ of order 20-30 GeV$^2$,
the asymptotic next-to-leading order formula (large $Q^2$) for the inclusive 
structure function differs from the exact result for a factor of 5\%
or less, indicating that at these not so large $Q^2$ values
the charm quark behaves already very much like a parton.
Since, as mentioned before, the boson-gluon fusion almost correctly
accounts for experimental results on single-inclusive distributions,
this points out that the scale at which the charm quark can be regarded
as a parton actually depends upon the physical observables considered.

Open heavy flavour photoproduction has been considered in 
ref.~\cite{Frixione_proc}. With an integrated luminosity 
of 1000~$pb^{-1}$, next-to-leading order QCD predicts
about $10^9$ ($10^6$) charm (bottom) particles produced
in $ep$ collisions (Weizs\"acker-Williams approximation).
Therefore, taking into account the experimental efficiencies, the 
number of reconstructed particles will be comparable to or larger than
the one obtainable by the fixed-target hadroproduction experiments
of the new generation at Fermilab. This will provide the HERA
experiments with the possibility of performing studies of
charm and bottom physics at an excellent level of accuracy.
In this respect, it is extremely interesting to compare the
results at HERA with the results of fixed-target photoproduction 
experiments, which have a center-of-mass energy of about
one order of magnitude smaller. In charm physics, high luminosity 
will allow to consider exclusive quantities, like correlations, which
constitute the most stringent test for the underlying theoretical
picture. One may also adopt a different point of view, namely
to look at charm production as a useful tool to constrain
the input parameters entering the calculations, like for
example the quark mass. To this end, it is mandatory to have
data with large statistics. The capability of HERA of producing
bottom quarks is also very promising. In particular, the comparison
of the QCD prediction for the $p_{\sss T}$ spectrum of bottom
(which has been shown in ref.~\cite{Frixione95b} to be only
marginally affected by the uncertainties on the input parameters of
the calculation) with the data could be of great help in understanding
the origin of the discrepancy observed at the Tevatron for the
same quantity.

\subsection{Production at HERA-B}

The study of the production mechanism of heavy flavours at HERA-B
might prove to be extremely useful. The data on bottom production
at fixed target have low statistics, and have been mainly
obtained in $\pi N$ collisions (only very recently, the first
measurement of the total $b\bar{b}$ rate in $pN$ collisions has been 
presented). A study of distributions would be interesting in order
to test the QCD description of bottom quark production at low
center-of-mass energies. In charm physics, the results of 
fixed-target experiments at CERN and Fermilab still leave plenty of 
open questions (see ref.~\cite{Frixione94} and references therein). 
In particular, new measurements may help in understanding 
the importance and the nature of non-perturbative
contributions to charm cross section. The hypothesis of the
intrinsic charm in the proton could be tested as well.

\subsection{Determination of the Gluon Density in the Proton 
using Charm Data}

The possibility has been considered of using charm
data to constrain (or to measure) the gluon density in
the proton, both in the unpolarized and the polarized case.
Photoproduction of charm is in principle an ideal tool in order
to perform this measurement, since the gluon density enters the
cross section already at the leading order, and in a simpler
way with respect to hadroproduction processes.
It has been shown in ref.~\cite{Frixione93a} that the use of correlations
between the charm and the anticharm can give a direct measure
of the unpolarized gluon density in the proton in a fairly large 
range in $x$ ($10^{-3}\div 10^{-1}$), since the contribution of the 
resolved component can be suitably suppressed. Being necessary
to reconstruct both the charm and the anticharm for this kind of
measurements, a very large integrated luminosity is required.
The measurement of the polarized gluon density in the proton
has been considered in refs.~\cite{Frixione96b,Stratmann96} 
(see also ref.~\cite{Polarized_ep}), in view of 
the possibility for HERA to operate in the polarized mode.
The conclusion has been reached~\cite{Frixione96b} that charm 
photoproduction data can be used to constrain this quantity, if an 
integrated luminosity of at least 100~$pb^{-1}$ will be achieved.

\subsection{Intrinsic Charm}

The possibility of detecting signals due to intrinsic charm in 
the proton has been discussed during the workshop by G.~Ingelman.
An intrinsic charm contribution has been suggested long time
ago~\cite{int_charm} to explain an excess of data from fixed-target
experiments with respect to theoretical predictions in the 
large-$x_{\sss F}$ region. Recently, this problem has been tackled
in ref.~\cite{Ingleman96}, dealing directly with $ep$ collisions
at HERA. It turns out that the intrinsic charm contribution
is non-negligible only in the very forward region, and can not be
detected with the present experimental configuration;
an upgrade in the very forward region would be necessary.

\section{Heavy Quark Production Cross Sections and Reconstruction
Efficiencies}
 
In this section we give measured cross sections and expected reconstruction
efficiencies for charmed mesons. As an example the capabilities
of the H1 detector~\cite{H1-detector} are given. The effect of a double
layer silicon vertex detector~\cite{H1-vertex} is investigated.

The charm production cross section has been measured
by both collaborations H1 and ZEUS~\cite{H1,Zeus} and is close to one
microbarn. The value for photoproduction ($Q^2 <0.01$~GeV$^2$) amounts
to~\cite{H1} 
\begin{displaymath}
\sigma(ep \rightarrow c\bar{c}X)=941 \pm  160 ^{+142}_{-120} ~nb.
\end{displaymath}
The first error is the statistical and the second the systematic
error dominated by the extrapolation uncertainty to the unmeasured
phase space. This rather large cross section makes HERA an ideal place
for charm physics.

\noindent The $b$ cross section has not been measured yet. 
An estimate~\cite{Eichler}  $\sigma(ep \rightarrow b \bar{b}X)=6 \pm  2 ~nb$
is of similar magnitude as on the Z$^0$- or  the $\Upsilon$-resonance and
much lower than at hadron colliders. 

The reconstruction of charm uses the central tracking chambers
with an angular coverage of $-1.5<\eta <1.5$. The decay chain
\begin{displaymath}
ep  \rightarrow D^*X
\rightarrow D^0 \pi_s \rightarrow (K^-\pi^+) \pi_s 
\end{displaymath}

\noindent yields a signal to background ratio of 1:1 for the mass
difference $\Delta M=M(K^-\pi^+ \pi^+)-M(K^-\pi^+)$ with a cut on the
transverse momentum of $p_{\sss T}(D^*)>2.5~GeV$.  A total efficiency in
photoproduction of trigger, acceptance, branching fraction, and
reconstruction of charmed events via $D^*$-tagging of 10$^{-4}$ has been
achieved (table \ref{Table:rate}).  This is only a tiny fraction of the
total cross section. The yield can be enlarged by  summing other decay
channels of the $D$ meson. A similar signal to background ratio can be
achieved with the reconstruction of  
\begin{displaymath}
D^0 \rightarrow
K^-\pi^+~(3.83\%),~K^0 \pi^+ \pi^- ~(5.4\%),  ~K^- \pi^+ \pi^+
\pi^-~(7.5\%),~K^- \mu^+ \nu~(3.23\%). 
\end{displaymath}
This is an improvement of a factor of 5 over the single
$K^- \pi^+$ channel. 

Besides the branching fractions into a
specific decay channel the biggest event loss (order of magnitude)
comes from the transverse momentum cut of the $D^*$ which is necessary
to reduce the combinatorial background. This can be partially
compensated with the help of a vertex detector.

\subsection{Charm Reconstruction with the H1 Vertex Detector} 

The main effect of a vertex detector is the assignment of charged
particle tracks to a secondary vertex which results in a much improved
signal to background ratio S/N, but with a corresponding loss of
acceptance. 

\begin{figure}[hbt]
\begin{center}
\epsfig{file= 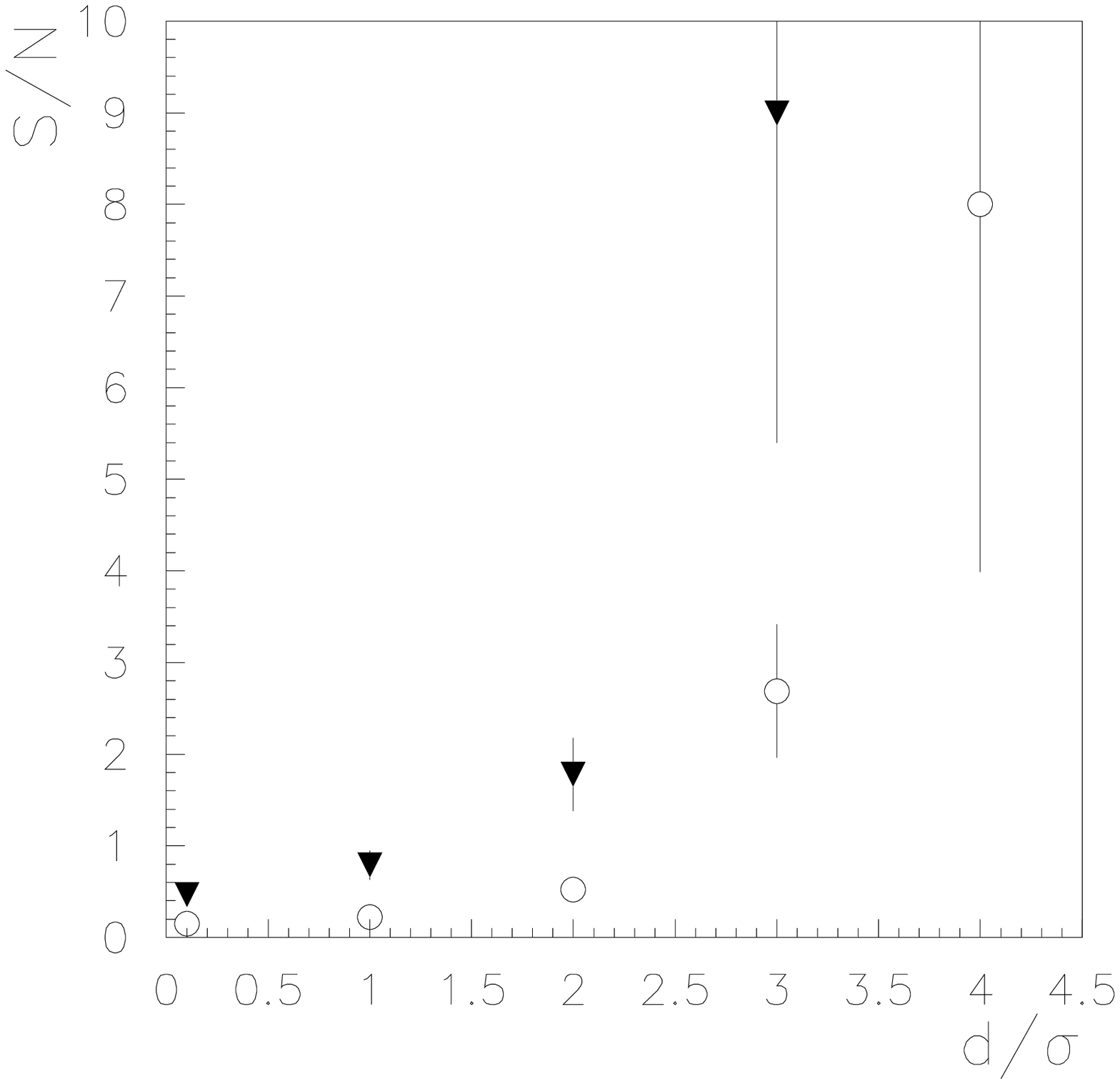, height=6.2 cm}
\epsfig{file= 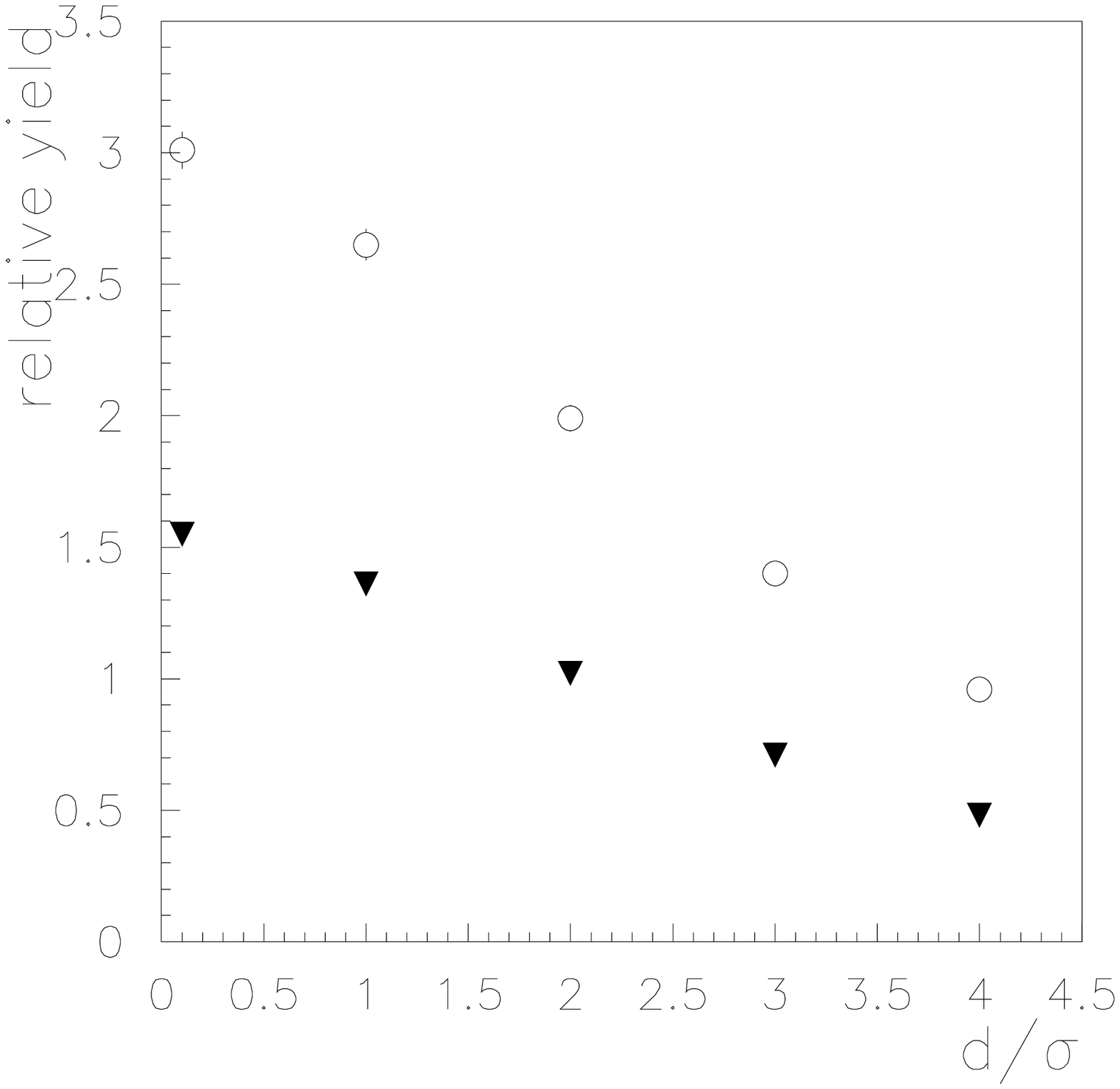 ,height=6.2 cm}
\put(-303.,150.){(a)}
\put(-130.,150.){(b)}
\vspace{1cm}
\caption[~]{\it \label{S/N} a): Signal to background
ratio S/N of the $D^*$ signal  versus the vertex separation $d$
normalized to its error $\sigma$ for a transverse momentum cutoff of 1
GeV (open circles) and 2 GeV (closed triangles) of the $D^*$. (b):  
Yield of the $D^*$ signal with $p_{\sss T}(D^*)>$1~GeV (open circles) 
and $>$2~GeV (closed triangles) respectively relative to a momentum cutoff of
$p_{\sss T}(D^*)>$2.5~GeV.}         
\end{center}
\end{figure}

This can be seen in figure \ref{S/N}a (from ref.~\cite{Prell}), 
where this ratio S/N is plotted versus the separation
$d/\sigma$ of primary and secondary vertex, with $d$  the distance
and $\sigma$ its error. No lower momentum cut for the $D^*$ was applied. 
The track finding efficiency of the slow pion in the $D^* \rightarrow
D^0 \pi_s$ decay is non zero only above $p_{\sss T}(\pi_s)$=120 MeV which 
implies an indirect  transverse momentum cutoff for the $D^*$ meson of 
about 1~GeV.

\normalsize The signal to background grows rapidly with the vertex
separation and reaches values above one for $d/\sigma>2-2.5$. At the
same time we observe a loss of the signal by a factor two compared to no
vertex cut (see figure \ref{S/N}b). Therefore the net gain is not very
large.

\begin{table}[htb]
\begin{center}
\begin{tabular}{|l | c c c c c|}
\hline
 {\bf no vertex}& acceptance& reconstr.& $c\rightarrow D^*
\rightarrow $&$D^0$ &total\\ 
{\bf  detector}&$p_{\sss T}>2.5$~GeV &efficiency&$D^0\pi$ 
& decay mode&efficiency  \\
\hline  $D^0 \rightarrow K^-\pi^+$&    0.03  & 0.5  &0.18 & 0.038  &1 10$^{-4}$
\\   $D^0 \rightarrow K^-\pi^+ \pi^+ \pi^-$  &     &      &     & 0.075  &  
\\  $D^0 \rightarrow K^-\mu^+ \nu$      &     &      &     & 0.032  &   \\ 
$D^0 \rightarrow K^0 \pi^+ \pi^-\rightarrow 4 \pi$   &     &      &     &
0.019  &   \\ \hline
 sum ($D^0$)               & 0.03 & 0.5   & 0.18 & 0.164 
&4~10$^{-4}$\\ \hline  
\hline
{\bf with vertex} & acceptance& reconstr.&
 $c\rightarrow  D^0/D^+$ & $D^0/D^+$ &total\\
{\bf detector} &$p_{\sss T}>1.0$~GeV &efficiency & &decay mode&efficiency  \\ 
& $d/\sigma>2.3$ & & & & \\ \hline
$D^0 \rightarrow K^-\pi^+  $& 0.05   & 0.3  &0.55 & 0.038  &3~10$^{-4}$ \\
$D^0 \rightarrow K^-\pi^+ \pi^+ \pi^-$  &     &      &     & 0.075  &   \\ 
$D^0 \rightarrow K^-\mu^+ \nu$      &     &      &     & 0.032  &   \\ 
$D^0 \rightarrow K^0 \pi^+ \pi^-\rightarrow 4\pi$   &     &      &     &
0.019  &   \\  \hline
      sum ($D^0$)  & 0.05 & 0.3   & 0.55 & 0.164   &1~10$^{-3}$\\ \hline
$D^+ \rightarrow K^0\pi^+ \rightarrow 3\pi$  &0.05 & 0.3 & 0.21&
0.019  &6~10$^{-5}$\\ 
$D^+ \rightarrow K^-\pi^+ \pi^+$      &     &     &      & 0.091  &   \\ 
$D^+ \rightarrow K^0 \pi^+ \pi^+ \pi^-\rightarrow 5 \pi$&    &     
&      & 0.024  &   \\ \hline
      sum ($D^{\pm}$)   & 0.05 & 0.3   & 0.21 & 0.134 &4~10$^{-4}$\\
\hline 
      sum (all $D$-mesons)&      &        &     &        &2~10$^{-3}$\\
\hline 
\end{tabular}
\caption[~]{\it \label{Table:rate} Summary of various reconstruction
efficiencies for $D^0$ mesons in photoproduction with and without a vertex
detector. Charged particles have been confined to $|\eta|<$1.5, and a vertex
separation of $d/\sigma>$2.3 has been assumed for the lower example with a
vertex detector. Acceptance and reconstruction efficiencies depend on
details of cuts and only rough numbers using cuts similar to ref.~\cite{H1}
are given.}      
\end{center}
\end{table}

\begin{table}[htb]
\begin{center}
\begin{tabular}{|l | c c|}
\hline
tagging method   & charm-tagging & bottom-tagging \\ \hline
$D^*$-meson      & 0.014 & 0.09 \\
identified muon  & 0.073 & 0.20 \\
non-vertex track, $d/\sigma>3$& 0.10 & 0.33 \\
non-vertex track, $d/\sigma>2$& 0.37 & 0.17 \\
\hline
\end{tabular}
\caption[~]{\it \label{Table:jets} Summary of charm and bottom tagging
efficiency inside jets with $E_{\sss T}^{jet} \ge 6~GeV$, $|\eta^{jet}|<2~$,
Q$^2<$4 GeV$^2$, and 135 GeV $<W_{\gamma p}<$ 270 GeV. Non-vertex track means
a track with an impact parameter $d$ larger than 2 resp. 3 times its error
$\sigma$ (from ref. \cite{Butterworth}).}       \end{center}
\end{table}

But with the vertex cut even  more decay channels are reconstructable.
The $D^0$ mesons can be found directly without the detour via the $D^*$.
This gains a factor of more than two. Also the $D^+$ decays into $K^-\pi^+
\pi^+ $(9.1\%), $K^0 \pi^+ \pi^+ \pi^-$ (7\%), and $K^0 \pi^+$ (2.8\%)
will be possible.  Therefore the net gain with a vertex detector to fully
reconstruct charmed final states is a factor 5-10.

A vertex detector is also useful for tagging charm by looking for
secondary vertices inside jets. A study~\cite{Butterworth} has been
performed where two or more jets with $E_{\sss T}^{jet} \ge 6~GeV$,
$|\eta^{jet}|<2~$, Q$^2<$4 GeV$^2$, and 135 GeV $<W_{\gamma p}<$ 270 GeV
were selected. The efficiency using  very loose cuts (at least one track
with an impact parameter displaced by d/$\sigma >$2) is 37\% for charm and
17\% for bottom with S/N=0.9 and also here  the gain from a vertex detector is
significant (order of magnitude).

Tables \ref{Table:rate} and \ref{Table:jets}
summarize the efficiencies.
For an integrated luminosity of 300~$pb^{-1}$ we expect therefore of order
10$^6$ fully reconstructed $D^0$  and $D^{\pm}$ mesons with a S/N$\ge$1.

\subsection{Charm-Anticharm Correlations}

For many applications and QCD studies the two-particle inclusive
distribution is of interest. We investigated the probability to fully
reconstruct two charmed particles in H1.

The acceptance that both charm quarks fall inside the acceptance of the
tracking system ($|\eta|<1.5$, $p_{\sss T}>1$ GeV) is 3\%. 
Using all the above decay
channels and a vertex detector we can reach a double tagging efficiency of
order  2.5$\times$10$^{-5}$, which translates into 7000 charm pairs for 
300~$pb^{-1}$. Without a vertex detector this number will be an order 
of magnitude less.

If we do not require the full reconstruction of both charmed mesons, but ask
only for one reconstructed $D$ meson plus a lepton of momentum 
$p_{\sss T}>1.5$ GeV
which does not come from the primary vertex, we reach similar efficiencies as
above, therefore doubling the number of double charm tags. 
Relaxing the quality of
the double tagged events even more one would ask for one fully reconstructed
charmed meson and a well separated  jet. Efficiencies depend very much on
chosen cuts and no numbers are given here.

\section{Summary and Comparison with other Machines}

At HERA, a rich physics program can be achieved with an integrated
luminosity ${\cal L}=$~300~$pb^{-1}$. The
inclusive properties of open charm photo- and electroproduction can be 
studied at an excellent level of accuracy. The next-to-leading order QCD
predictions for the direct color-singlet photoproduction of $J/\psi$
can be thoroughly tested. Correlations between charm and anticharm,
open bottom production, resolved and direct color-octet quarkonium 
photoproduction could be studied with ${\cal L}=$~300~$pb^{-1}$ as well,
although larger luminosity would allow a more significant comparison
between theory and experiments. Charm correlations can be exploited in
order to measure the gluon density in the proton. In polarized $ep$
collisions, charm data are useful to constrain the {\it polarized} gluon 
density in the proton. Very large luminosity is mandatory
in order to study the production of $\chi$, $J/\psi +\gamma$ and 
to investigate the charm fragmentation into $J/\psi$. Intrinsic charm
signals are not detectable without upgrading the detectors in
the very forward region. A vertex detector is essential to reach a 
high tagging efficiency for open charm and should be taken into account 
in the design of the luminosity upgrade program.

A figure of merit of the potential for charm physics is the total number of
reconstructed open charmed particles.  The biggest competition comes from fixed
target experiments at Fermilab. A program to reach 10$^6$ reconstructed open
charm is in place which corresponds to 300 $pb^{-1}$ at HERA. Progress beyond
10$^6$ will depend on commitment to the physics and a potential to 
get 10$^8$ at Fermilab is enticing~\cite{Appel}. 

\vspace*{0.5cm}
\noindent {\bf Acknowledgements:} One of the authors (S.F.) wishes
to thank the Swiss National Foundation for financial support.

\end{document}